\documentclass[12pt]{article}

\usepackage{graphics} 
\usepackage{blindtext}
\usepackage{epsfig} 
\usepackage{subfigure}
\usepackage{amsmath} 
\usepackage{amssymb}  
\usepackage{color}
\usepackage{lipsum}
\usepackage{cuted}
\usepackage{enumerate}

\title{\LARGE \bf
Identification of Switched Autoregressive Systems \\ from Large Noisy Data Sets
}

\author{Sarah Hojjatinia$^{1}$,  Constantino M. Lagoa$^{2}$, and Fabrizio Dabbene$^{3}$ 
\thanks{$^{1}$Sarah Hojjatinia is with the School of Electrical Engineering and Computer Science,
        The Pennsylvania State University, University Park, PA, USA,
        {\tt\small szh199@psu.edu}}%
    \thanks{$^{2}$Constantino M. Lagoa is with  the School of Electrical Engineering and Computer Science,
    	The Pennsylvania State University, University Park, PA, USA,
    	{\tt\small lagoa@psu.edu}}%
    \thanks{$^{3}$ Fabrizio Dabbene is with CNR-IEIIT, Politecnico di Torino, 10129 Torino, Italy, {\tt\small fabrizio.dabbene@ieiit.cnr.it}\newline
\indent
This work was partially supported by National Institutes of Health (NIH) Grant R01 HL142732, National Science Foundation (NSF) Grant \#1808266 and the 
International Bilateral Joint CNR-JST Lab COOPS.}
}

\newtheorem{assumption}{Assumption} 
\newtheorem{problem}{Problem} 
\newtheorem{thm}{Theorem}
\newtheorem{lemma}{Lemma}

\date{}
\begin{document}

\maketitle
\thispagestyle{empty}
\pagestyle{empty}

\begin{abstract}
The paper introduces a novel methodology for the identification of coefficients of switched autoregressive   linear models. We consider the case when the system's outputs are  contaminated by possibly large values of measurement noise. It is assumed that only partial information on the probability distribution of the noise is available. Given input-output data, 
we aim at identifying switched system coefficients  and parameters of the distribution of the noise	which are compatible with the collected data.  
System dynamics are estimated through expected values  computation and by exploiting the  strong law of large numbers. 
We demonstrate the efficiency of the proposed approach with several academic examples. 
The method is shown to be extremely effective in the situations where a large number of measurements is available; cases in which previous approaches based on polynomial or mixed-integer optimization  cannot be applied due to very large computational burden.
\end{abstract}

\section{Introduction}

The interest in the study of hybrid systems has been persistently growing in the last years, due to their capability of describing real-world  processes in which continuous and discrete time dynamics coexist and interact. Besides classical automotive and chemical processes, emerging applications include computer vision, biological systems, and communication networks.

Moreover, hybrid systems can be used to efficiently approximate nonlinear dynamics,
with broad application, ranging from civil structures to robotics and systems biology,
that entail extracting information from high volume data streams
\cite{ozay2014convex},  \cite{sznaier2014surviving}.
In the case of high dimentional data, nonlinear order reduction or  low dimensional sparse representations techniques  \cite{lin2006learning}, \cite{hojjatinia2018parsimonious}, \cite{saul2003think}, are very effective in handling static data, but most do not exploit
dynamical information of the data.

In the literature, several results have been obtained for the analysis and control  of hybrid systems, formally characterizing important properties such as stability or reachability, and proposing different control designs \cite{lunze2009handbook}.
In parallel,  researchers rapidly realized that first-principle models may be hard to derive especially with the increase of diverse application fields. This sparked interest on the problem of identifying hybrid (switched) models starting  from experimental data; see for instance the tutorial paper \cite{paoletti2007identification}
and the survey \cite{garulli2012survey}.

It should be immediately pointed out that this identification problem is not a simple one, since the simultaneous presence of  continuous and discrete state variables gives it a combinatorial nature. The situation becomes further complicated in the presence of unknown-but-bounded noise. In this case the problem is in general NP-hard.  
Several approaches have been proposed to address this difficulty, see e.g.~\cite{lauer2011continuous}.
The paper \cite{roll2004identification}
reformulates the problem as a mixed-integer program.
These techniques proved to be very effective in situations involving relatively small noise levels or moderate dimensions,
but they do not appear to scale well, and 
their performance deteriorates as the noise level or problem size increase.

Of particular interest are recent approaches based on convex optimization: in \cite{Bako:2011:ISL:1963659.1963792}
some relaxation based on sparsity are proposed, while \cite{Ozay2015180} develops a moment based approach to identify the switched autoregressive exogenous system, and \cite{HOJJATINIA201714088} adapts it toward Markovian jump systems identification. 
These methods are surely more robust, and represent the choice of reference for medium-size problems
and medium values of noise, and have found applications in several contexts, ranging from segmentation problems arising in computer vision to biomedical systems. 

However, the methods still rely on the solution of rather large optimization problems. Even if the convex nature of these problems allows to limit the complexity growth, there are several situations for which their application becomes critical. For instance, identification problems cases that involve quite high  noise levels and/or large number of measurements. 

An enlightening example, which serves as a practical motivation for our developments, arises in healthcare applications: the availability of  \textit{activity tracking devices} allows to gather a large amount of information of the physical activity of an individual. Physical activity is a dynamic behavior, which in principle can be modeled as a dynamical system \cite{Lagoa2017}. Moreover,  its characteristics may significantly change depending on the time of the day, position, etc. This motivated the approach of modeling it as a switching system  \cite{conroy2019}.

In this paper, we focus on cases involving a very large number of sample points, possibly affected by large levels of noise. In this situation, polynomial/moments based approaches become ineffective, and different methodologies need to be devised.
The approach we propose builds upon the same premises as  \cite{Ozay2015180}:
the starting point is the algebraic procedure due to Ma and Vidal \cite{Ma2005},
where it has been shown for noiseless processes, it is possible to identify the different subsystems in a switching system by recurring to a Generalized Principal Component Analysis (GPCA).
In particular,  we infer the parameters of each subsystem from the null space of a matrix $V_n(r)$ constructed from the input-output data $r$ via a nonlinear embedding (the Veronese map). 

The approach was extended to the cases in presence of process noise in \cite{Ozay2015180}, showing how the  entries of this matrix depend polynomially on the unknown noise terms. Then, the problem was formulated in an unknown-but-bounded setting, looking for  an admissible noise sequence rendering the matrix $V_n(r)$ rank deficient. This problem was then relaxed using polynomial optimization methods.

In this work, we follow the same line of reasoning, but then take a somewhat different route. First, we consider random noise, and we assume  that \textit{some} information on the noise is available. 
Then, instead of relaxing the problem, we exploit the availability of a large number of measurements to make recurse to law-of-large-numbers type of reasoning. This allows us to devise an algorithm characterized by an extremely low complexity in terms of required operations. 
The ensuing optimization problem involves only the computation of the singular vector associated with the minimum singular value of a matrix that can be efficiently computed and whose size does not depend on the number of measurements.

 \subsection{Paper Organization}

 
The paper is structured as follows:   
after this introduction, there is a brief notation section. Section \ref{stat} includes the problem statement. In Section \ref{review}, 
algebraic reformulation of  switched autoregressive  (SAR) linear system identification problem for noiseless data is reviewed. The problem of identifying SAR system in the presence of noise is surveyed in Section \ref{noisy}.
In Section \ref{estim}, the algorithm for estimating unknown noise parameters is described. Numerical results are shown in Section \ref{result}.
Finally, Section \ref{conclusion} concludes the paper highlighting some possible future research directions.

\subsection{Notation} \label{sec:notation}
Given a scalar random variable $X$, we denote by $m_{d}$ its $d^{th}$ moment, which  may be computed according to the following integral 
\begin{equation} \label{moment}
m_{d}=E[x^{d}]=\int_{-\infty}^{\infty}x^{d}\, f(x)\, dx
\end{equation}
where $E[\cdot]$ refers to expectation, and  $f(x)$ is the probability density function of $X$. Additionally, the  variance of $X$ is indicated by $s^2$.
For instance, if $X$ has a normal distribution with zero mean and variance $s^2$, i.e. $f(x)=\dfrac{1}{s\, \sqrt{2\pi}}e^{-x^{2}/2s^{2}}$, its
moments are given by
\begin{equation}
m_{d}=E[x^{d}]=\begin{cases}
0 ~& \text{if}~ d~ \text{is odd}\\ 
s^{d} \, (d-1)!! ~& \text{if} ~d~ \text{is even}
\label{eq:mn}
\end{cases}
\end{equation}
where $!!$ denotes  double factorial ($n!!$ is  the product of all numbers from $n$ to 1 that have the same parity as $n$).


 \section{Problem Statement}\label{stat}

In this section,  a complete description of the problem is addressed. In addition, the required assumptions are defined to solve the problem.

 \subsection{System Model}
We consider  SAR systems  of the form
  \begin{align}
  x_{k}=\sum_{j=1}^{n_{a}}{a_{j  \sigma(k)}} \; x_{k-j}+\sum_{j=1}^{n_{c}}{c_{j\sigma(k)}} \;u_{k-j}    \label{eq:C}
  \end{align}
where $x_{k} \in \mathbb{R}$ is the output at time $k$ and $u_{k} \in \mathbb{R}$ is input at time $k$. The variable $\sigma(k) \in \{1, . . . , n\}$ denotes the sub-system active at time $k$, where  $n$ is the total number of sub-systems. Furthermore,  $a_{j  \sigma(k)}$ and $c_{j  \sigma(k)}$ denote unknown coefficients corresponding to mode $\sigma(k)$. Time $k$ takes values over the non-negative integers.

In practice,  output is always contaminated by noise; i.e. we assume that we observe 
  \begin{align}
  y_{k}= x_{k}+\eta_{k}     \label{eq:C1}
  \end{align}
  where $\eta_{k}$, denotes measurement noise. 

The following assumptions are made on the system model and noise.

\begin{assumption} \label{ass} 
Throughout this paper it is assumed that:
\begin{itemize} 
	\item  Upper bounds on $n_a$ and $n_c$ are available.
	\item  Upper bound on the number of subsystems $n$ is available.
	\item  Noise $\eta_{k}$ at time $k$ is independent from $\eta_{l}$ for $k\neq l$, and identically distributed with probability density
	$
	f(\eta|\theta)
	$;
	where $\theta$ is a (low dimensional) vector of unknown  parameters.
	\item  Input sequence $u_{k}$  applied to the system is known and bounded. 
	\item  There exists a finite constant $L$ so that  $|x_{k}|\leq L$ for all positive integers $k$. 
\end{itemize}  
\end{assumption}  
    
\subsection{Problem Definition}

The main objective of this paper is to develop algorithms that are able to identify the coefficients of a SAR model from noisy observations. More precisely, we aim at solving the following problem:

\begin{problem}
Given Assumption~\ref{ass}, an input sequence $u_k$, $k=-n_c+1, \dots, N-1$ and noisy output measurements $y_k$, $k=-n_a+1, \dots, N$,
determine coefficients of the SAR model $a_{i,j}$, $i=1,2,\ldots, n_a$, $j=1,2,\ldots,n$, $c_{i,j}$, $i=1,2,\ldots, n_c$, $j=1,2,\ldots,n$,
and the noise distribution parameters~$\theta$.\end{problem}

\section{Noiseless Case: A Review} \label{review}

As a motivation for the approach presented in this paper, we review and slightly reformulate earlier results on an algebraic reformulation of  the SAR identification problem for the case where no noise is present. We refer the reader  to~\cite{VidalSoattoMaEtAl2003}
for 
details on this algebraic approach to switched system identification.

\subsection{ Hybrid Decoupling Constraint} \label{hybrid}
We start by noting that equation~(\ref{eq:C}) is equivalent to
\begin{align}
b_{\sigma(k)}^{T} \; r_{k}=0  \label{eq:D}
\end{align}
where we introduced the (known) regressor at time $k$ 
\[
r_{k}=[x_{k},~ x_{k-1}, ~\cdots, ~x_{k-n_{a}}, ~u_{k-1}, ~\cdots, ~u_{k-{n_{c}}}]^{T}
\]
and the vector of unknown coefficients at time $k$
\begin{multline*}
{b_{\sigma(k)}=} \\~ [-1,~a_{1 \sigma(k)}, ~\cdots, ~a_{n_{a} \sigma(k)},~c_{1 \sigma(k)}, ~\cdots, ~c_{n_{c} \sigma(k)}]^{T}.
\end{multline*}
Hence, independently of which of the~$n$ submodels is active at time $k$, we have
\begin{align}
P_{n}(r_{k})= \prod_{i=1}^{n} {b_{i}^{T} r_{k}}=c_{n}^{T} \nu_{n}(r_{k})=0,  \label{eq:E}
\end{align}
where the
vector of parameters corresponding to the $i$-th submodel is denoted by $b_{i} \in \mathbb{R}^{n_{a}+n_{c}+1}$,  and $\nu_{n}(�)$ is  Veronese map of degree $n$ \cite{harris2013algebraic} 
\[
\nu_{n}([x_{1},~ \cdots,~  x_{s}]^{T} ) = [\cdots,~ x_{1}^{n_{1}}x_{2}^{n_{2}} \cdots x_{s}^{n_{s}} ,~\cdots]^{T}
\]
which contains all monomials of order $n$ in lexicographical order,
and $c_{n}$ is a vector whose entries are polynomial functions of unknown parameters $b_{i}$  (see \cite{vidal2005generalized} 
for explicit definition). 
The Veronese map above is also known as  polynomial embedding in machine learning \cite{vidal2005generalized}. 
%

Equation \eqref{eq:E} holds for all $k$, and  these equalities  can be expressed in matrix form
\begin{align}
V_{n}(r)c_{n}=\left[
\begin{matrix}
\nu_{n}(r_{1})^{T},~ \cdots,~ \nu_{n}(r_{N})^{T}
\end{matrix}
\right ]^{T} c_{n}=0   \label{eq:F}
\end{align}   
where $r$, without the subscript, denotes the set of all regressor vectors. 
Clearly, we are able to identify  $c_{n}$ (and hence the system's parameters) if and only if $V_{n}(r)$ is rank deficient. In that case, the vector $c_{n}$ can
be found by computing the nullspace of $V_{n}$.
\subsection{A Reformulation of the Hybrid Decoupling Constraint } 

Note that the number of rows of the Veronese matrix $V_n$ is equal to the number of measurements available for the regressor; i.e., in the notation of our paper, the number of rows is $N$. Therefore, a reformulation of the results in the previous section is needed to be able to address the problem of identification from very large data sets.

As mentioned in the previous section, in the absence of noise, the SAR system identification is equivalent to finding a vector $c_n$ satisfying
\[
c_{n}^{T} \nu_{n}(r_{k})=0 \text{ for all } k=1,2,\ldots N. 
\]
This is in turn equivalent to finding $c_n$ so that
\[
\frac{1}{N} \sum_{k=1}^N c_{n}^{T} \nu_{n}(r_{k}) \nu_{n}^T(r_{k})c_{n} = 0
\]
As a result, for the noiseless case, identifying the coefficients of the sub-models is equivalent to finding the singular vector $c_n$ associated with the minimum singular value of the matrix
\begin{equation} \label{eq:Mk}
\overline{\mathcal{M}}_N=\frac{1}{N} \sum_{k=1}^N \nu_{n}(r_{k}) \nu_{n}^T(r_{k}) \doteq \frac{1}{N} \sum_{k=1}^N M_k
\end{equation}
Note that, by using this equivalent condition, we only need to consider matrices of size  $\binom{n+n_{a}+n_{c}}{n}$. In other words, the size of this matrix \emph{does not depend} on the number of measurements. This is especially important when considering very large data sets.

\section{SAR system Identification in the Presence of Noise} \label{noisy}

Now, we address the case where the measurements of  output of the  switched autoregressive system are corrupted by noise. As a first step, we consider the case where the distribution of the noise is known, so  its moments $m_d$ are available. 

As seen in the previous section, identifying the parameters of the SAR model is equivalent to finding a vector in the null space of the matrix
$\overline{\mathcal{M}}_N$.
Under mild conditions, the null space of the matrix above has dimension one if and only if the data is compatible with the assumed model. However, if noise is present, $x_k$ is not known; therefore, this matrix cannot be  computed. In this section, we use available information on the statistics of the noise to compute approximations of the matrix $\overline{\mathcal{M}}_N$,  consequently approximations of vectors in its null space.

\subsection{On the Powers of $x_k$}
Since we do not have access to the values of the output $x_k$ to estimate the values of the quantities in equation~\eqref{eq:Mk}, we need to relate the powers of $x_k$ to the measurements and 
available information of the noise.

Note that $x_{k}$ is a (unknown) deterministic quantity. Therefore
\begin{equation} 
x_{k}^{h}=E[x_{k}^{h}]
\end{equation}
Since $x_{k}=y_{k}-\eta_{k}$ we have 
\begin{equation} 
\begin{aligned}
x_{k}^{h}=E[x_{k}^{h}]=E[(y_{k}-\eta_{k})^{h}]
~~~\forall k=1, \, 2,\, \cdots, \, N
\end{aligned}
\end{equation}
Assume, for simplicity  the distribution of the noise is symmetric with respect to the origin. As a result, all odd moments are zero
(in particular, the noise is zero mean, i.e.\ $m_1=0$). This assumption is made to simplify the calculations below and the approach can be immediately extended to the non-symmetric case. 

We concentrate on computing the expected value of powers of $x_{k}$ recursively and in a closed form.
First, we give an example of how to compute the expected value of powers of $x_{k}$ for  powers  $h=1,2$.
For $h=1$, we have
\begin{equation}
x_{k}=E[x_{k}] =E[y_{k}-\eta_{k}]=E[y_{k}]-m_{1}=E[y_{k}]\\
\end{equation}
while, for $h=2$, we can write
\begin{equation} \label{16}
x_{k}^{2}=E[x_{k}^{2}] =E[(y_{k}-\eta_{k})^{2}]=E[y_{k}^{2}] - 2E[y_{k}\eta_{k}]+ E[\eta_{k}^{2}].
\end{equation}
Note 
 that  $E[y_{k}^{2}]$ can be estimated from collected data, and $E[\eta_{k}^{2}]$ is equal to  second moment of noise ($m_{2}$), which is assumed to be known. To estimate the value of $E[y_{k}\eta_{k}]$, consider the following
\begin{equation} \label{17}
E[y_{k}\eta_{k}]=E[(x_{k}+\eta_{k})\eta_{k}]=E[x_{k}\eta_{k}]+E[\eta_{k}^{2}].
\end{equation}
The quantities $x_{k}$ and $\eta_{k}$ are mutually independent and, therefore,  $E[x_{k}\eta_{k}]=E[x_{k}]E[\eta_{k}]$, with $E[\eta_{k}]=m_{1}=0$. As a consequence, we have
\begin{equation} \label{18}
E[y_{k}\eta_{k}]=E[\eta_{k}^{2}]
\end{equation}
and finally  the value of equation \eqref{16} is
\begin{align}
E[x_{k}^{2}] & =E[y_{k}^{2}] - 2E[\eta_{k}^{2}]+ E[\eta_{k}^{2}]
=E[y_{k}^{2}] -  E[\eta_{k}^{2}] \\ & =E[y_{k}^{2}] -  m_{2} \nonumber
\end{align}

\begin{figure*}[t]
	\setlength\stripsep{\partopsep}%
	\[
	M_{k}=\nu_{n}(r_{k}) \,\nu_{n}^{T}(r_{k})=\]\[
	\begin{pmatrix}
	x_{k}^4 & x_{k}^3\, x_{k-1}  & x_{k}^3\, u_{k-1}  &  x_{k}^2\,x_{k-1}^2 &  x_{k}^2\,x_{k-1}\,u_{k-1}&  x_{k}^2\,u_{k-1}^2 \\
	*  &        x_{k}^2\,x_{k-1}^2   &  x_{k}^2\,x_{k-1}\,u_{k-1} &   x_{k}\,x_{k-1}^3 & x_{k}\,x_{k-1}^2\, u_{k-1}  & x_{k}\,x_{k-1}\,u_{k-1}^2\\
	*    & *   &              x_{k}^2\,u_{k-1}^2    &       x_{k}\,x_{k-1}^2\,u_{k-1}        & x_{k}\,x_{k-1}\,u_{k-1}^2   &  x_{k}\,u_{k-1}^3 \\
	* &    *    &       *       & x_{k-1}^4  & x_{k-1}^3\,u_{k-1}  &  x_{k-1}^2\,u_{k-1}^2 \\
	*    &    *    &  *    &  *     &        x_{k-1}^2\,u_{k-1}^2   &  x_{k-1}\,u_{k-1}^3\\
	*   &  * &   *   &   *   &  *  &           u_{k-1}^4
	\end{pmatrix}
	\]
	%
	\newcommand\scalemath[2]{\scalebox{#1}{\mbox{\ensuremath{\displaystyle #2}}}}
	\setlength\stripsep{\partopsep}%
	%
	\begin{align*} M_k &=
	\scalemath{0.88}{
		\left(
		\begin{smallmatrix}
		E[y_{k}^4]-6 \, m_{2}\,(E[y_{k}^2]-m_{2})-m_{4} & (E[y_{k}^3]-3\, m_{2} \, E[y_{k}]) \, E[y_{k-1}]  & (E[y_{k}^3]-3\, m_{2 }\, E[y_{k}])\, u_{k-1}   \\
		* &        ~(E[y_{k}^2]-m_{2})\,(E[y_{k-1}^2]-m_{2})   &  ~(E[y_{k}^2]-m_{2})\,E[y_{k-1}]\,u_{k-1} \\
		*  &  *   &              (E[y_{k}^2]-m_{2})\,u_{k-1}^2  \\
		*   & *    &   * \\
		* & *  &*  \\
		*  & * & *
		\end{smallmatrix} 
		\right.
	} \cdots \\ \\ & \hskip 1.5in \cdots \scalemath{0.72}{
		\left.
		\begin{smallmatrix}
		(E[y_{k}^2]-m_{2})\,(E[y_{k-1}^2]-m_{2}) &  (E[y_{k}^2]-m_{2})\,E[y_{k-1}]\,u_{k-1}&  (E[y_{k}^2]-m_{2})\,u_{k-1}^2\\
		(E[y_{k-1}^3]-3\, m_{2}\,E[y_{k-1}])\,E[y_{k}] & (E[y_{k-1}^2]-m_{2})\,E[y_{k}]\, u_{k-1}  & E[y_{k}]\,E[y_{k-1}]\,u_{k-1}^2 \\
		(E[y_{k-1}^2]-m_{2})\,E[y_{k}]\,u_{k-1}       & E[y_{k}]\,E[y_{k-1}]\,u_{k-1}^2   &   E[y_{k}]\,u_{k-1}^3 \\
		~~E[y_{k-1}^4]-6\,m_{2}\,(y_{k-1}^2-m_{2})-m_{4}   & ~~(E[y_{k-1}^3]-3\,m_2\,E[y_{k-1}])\,u_{k-1} &  ~~(E[y_{k-1}^2]-m_2)\,u_{k-1}^2 \\
		*   &        (E[y_{k-1}^2]-m2)\,u_{k-1}^2  &  E[y_{k-1}]\,u_{k-1}^3 \\
		* & *&           u_{k-1}^4
		\end{smallmatrix}
		\right)}
	\end{align*}
	\caption{Example of construction of $M_k$} \label{fig:Mk}
\end{figure*}

The reasoning above can be generalized to any power of~$x_{k}$. More precisely, we have the following result whose  proof is an immediate consequence of the derivations so far.

\vskip .1in
\begin{lemma}
	The   expected value of  the powers of $x_{k}$ satisfies
	\begin{align} 
	E[x_{k}^{h}] &=E[(y_{k}-\eta_{k})^{h}] =E[y_{k}^{h}] - \sum_{d=1}^{h} \binom{h}{d}\, E[x_{k}^{h-d}]\, E[\eta_{k}^{d}] \nonumber \\
	&=E[y_{k}^{h}] - \sum_{d=1}^{h} \binom{h}{d}\, E[x_{k}^{h-d}]\, m_{d} \nonumber \\
	&\forall k=1, \, 2,\, \cdots, \, N. \label{eq:rec}
	\end{align} \vskip .2in
\end{lemma}

\subsection{On the Structure of $M_k$}

We derive  some of the properties of the matrices $M_k$  now. An immediate consequence of the results in the previous section is the following.

\vskip .1in 
\begin{lemma} \label{lemma:Mk}
	Assume that  the noise distribution, some parameters of the noise, and the input signal are given and fixed. Let $mon_n(\cdot)$ denote a function that returns a vector with all monomials up to order $n$ of its argument. Then there exists an affine function $M(\cdot)$ so that
	\begin{align*}
	M_k &= E\{M[mon_n(y_k,\ldots,y_{k-n_a})]\}\\& = M\{E[mon_n(y_k,\ldots,y_{k-n_a})]\}.
	\end{align*}
\end{lemma} \vskip .2in

It should be noted 
that the random variables $y_k$ and $y_l$ are mutually independent for $k\neq l$, so the function above can further be represented as a multilinear function of the moments of $y_k$.


\subsection{An Example of Construction of $M_k$}

To better illustrate the approach used in this paper, we provide an example of how to construct the  matrix $M_{k}$ required for identification. 
To this end, consider the problem of identifying a SAR system  with $n=2$ subsystems of the form
\begin{equation} \label{eq:ex}
\begin{aligned}
\text{subsystem 1}: ~~& x_{k}= a_{1} \, x_{k-1} +b_{1}\, u_{k-1} \\
\text{subsystem 2}:~~ & x_{k}= a_{2} \, x_{k-1} +b_{2}\, u_{k-1}  \\
\end{aligned}
\end{equation}
from noisy measurements
\begin{align} 
y_{k}= x_{k}+\eta_{k}  
\end{align}
where $\eta_{k}  $ has a symmetric distribution.
We can rewrite the system as in equation \eqref{eq:E}. In particular, the vector $c_2$ as  a function of the parameters of the subsystems, assumes the form
\begin{align*}
c_{2}&= [1,  -(a_{1}+a_{2}), -(b_{1}+b_{2}),  a_{1}a_{2},   a_{1}b_{2}+b_{1}a_{2},  b_{1}b_{2}
]^{T}.
\end{align*}
The regressor vector $r_{k}$ at time $k$ 
\begin{equation*} 
{r_{k}}=
\begin{bmatrix}
x_{k} &
x_{k-1} & 
u_{k-1} 
\end{bmatrix}^T
\end{equation*}
gives rise to the following Veronese vector  
\begin{equation} \label{eq:vn}
\begin{aligned}
\nu_{n}(r_{k}) = r_{k} \otimes r_{k} =
\begin{bmatrix}
x_{k}^{2} \\
x_{k} \, x_{k-1}\\ 
x_{k} \, u_{k-1}\\
x_{k-1}^{2} \\
x_{k-1} \, u_{k-1}\\ 
u_{k-1}^{2} \\
\end{bmatrix}
\end{aligned}
\end{equation}
\newline
whose size is $l\times 1$, with $l= \binom{n+n_{a}+n_{c}}{n}=\binom{2+1+1}{2}=6$.
From $r_{k}$ and $\nu_{n}(r_{k})$, we can  compute matrix $M_{k}$, which is given in Figure~\ref{fig:Mk}.
Then, as we have the values of noisy output $y_k$, we compute expected value of powers of $x_k$ in terms of expected value of powers of $y_k
$ and moments of measurement noise. Following the results of Lemma 1, we obtain the second matrix in Figure~\ref{fig:Mk}. 
For system of equation \eqref{eq:ex}, $M_{k}$ is given by the two expression in Figure~\ref{fig:Mk}.
%

\subsection{Identification Algorithm}

As mentioned before, to identify the parameters of the SAR system, we need to be able to estimate the matrix~$\overline{\mathcal{M}}_N$ in equation \eqref{eq:Mk}. It turns out that it can be done by using the available noisy measurements. More precisely, we have the following result.

\vskip .1in
\begin{thm} \label{thm1}
	Let $M(\cdot)$ and $mon_n(\cdot)$ be the functions defined in Lemma~\ref{lemma:Mk}. Define
	\[
	\widehat{\overline{\mathcal{M}}}_N \doteq \frac{1}{N} \sum_{k=1}^N M[mon_n(y_k, \ldots, y_{k-n_a})]
	\]
	Then, as  $ N\rightarrow \infty$,
	\[
	\widehat{\overline{\mathcal{M}}}_N - {\overline{\mathcal{M}}}_N \rightarrow 0 \text{ a.s.} 
	\] \vskip .2in
\end{thm}

\noindent
\textbf{Sketch of proof:} See Appendix.

As a result, the empirical average computed using the noisy measurements (where expected values of monomials are replaced by the measured monomial values) converges to the desired matrix in equation~\eqref{eq:Mk}. Therefore we propose the following algorithm for identification of a SAR system.

\vskip .1in 
\noindent
\emph{Algorithm 1:} Let $n_a$, $n_c$, $n$ and some parameters of the noise be given. 
\begin{enumerate}[Step 1.] 
	\item Compute matrix
	\[
	\widehat{\overline{\mathcal{M}}}_N=\frac{1}{N} \sum_{k=1}^N M[mon_n(y_k, \ldots, y_{k-n_a})]
	\]
	\item Let $c_n$ be the singular vector associated with the minimum singular value of $\widehat{\overline{\mathcal{M}}}_N$.
	\item Determine the coefficients of the subsystems from the vector~$c_n$.
\end{enumerate}

In order to perform Step 3 in Algorithm 1, we adopt 
polynomial differention algorithm for mixtures of hyperplanes, 
introduced by Vidal \cite[pp.~69--70]{vidal2003generalized}. 
\section{Estimating Unknown Noise Parameters } \label{estim}

We now address the case where the distribution of the noise is not completely known. As mentioned in Assumption~\ref{ass}, the distribition of the noise is known except for a few parameters $\theta$. 
For simplicity of exposition, lets consider the case where the noise has a normal distribution with zero mean and unknown variance $s^2$. The reasoning extends to any noise distribution with a small number of  unknown parameters.

In such a case, the objective is to simultaneously estimate  system parameters and the variance of  noise. We start by noting  that computing ${\overline{\mathcal{M}}}_N$ using the true value of the variance results in a rank deficient matrix. Moreover, given collected data $y_k$ and $u_k$, the matrix 
$\widehat{\overline{\mathcal{M}}}_N$
is a continuous function of the moments of  noise and, hence, a \emph{known} continuous function of the standard deviation~$s$. Given previous convergence results,  the true value of $s$ will make  $\widehat{\overline{\mathcal{M}}}_N$ to have a very small minimum singular value (especially for large values of $N$). For this reason,  estimation of $s$ can be performed by minimizing the minimum singular value of matrix above over the allowable values of $s$. More precisely, we propose the following algorithm

\vskip .1in 
\noindent
\emph{Algorithm 2:} Let $n_a$, $n_c$, $n$, some parameters of the noise  and $s_{\max}$  be given. 
\begin{enumerate}[Step 1.] 
	\item Compute matrix
	\[
	\widehat{\overline{\mathcal{M}}}_N = \frac{1}{N} \sum_{k=1}^N M[mon(y_k, \ldots, y_{k-n_a})]
	\]
	as a function of the noise parameter $s$.
	\item  Find the value  $s^* \in [0,s_{\max}]$ that minimizes the minimum singular value of $\widehat{\overline{\mathcal{M}}}_N$.
	\item Let $c_n$ be associated singular vector.
	\item Determine the coefficients of the subsystems from the vector~$c_n$.
\end{enumerate}

\vskip .1in 

Note that the nonconvex optimization of Step 2 can be solved via an easily implementable line-search. However, the solution $s^*$ might not be unique; i.e.,  there might exist several values of $s$ that lead to a minimum singular value very close to zero. 
In practice, our experience has been that, for sufficiently large $N$, the above algorithm provides both a good estimate of the systems coefficients, and noise parameters; especially if we take $s^*$ to be the smallest value of $s$ for which the minimum singular value of $\widehat{\overline{\mathcal{M}}}_N$ is below a given threshold $\epsilon.$


\begin{table*}[ht]
	\caption{Identifying polynomial coefficients for different values of noise variance and different system run.}
	\label{tab:table1}
	\centering
	\begin{tabular}{||l|c|c|c|c|c||} 
		\hline \hline
		\textbf{system} &\textbf{Value 1} & \textbf{Value 2} & \textbf{Value 3}& \textbf{Value 4} & \textbf{Value 5} \\
		\text{\#} &1 & $-(a_{1}+a_{2})$ & -$(b_{1}+b_{2})$ & $a_{1}\,a_{2}$ & $a_{1}\,b_{2}+b_{1}\,a_{2}$ \\
		\hline \hline
		\text{true parameters}	&1 & 0.2 &0 & -0.15 & -0.8  \\
		\hline\hline
		\text{identification 1}	&1	&  0.2002 &  0.0001 &  -0.1503 &  -0.7989 \\
		\hline
		\text{identification 2}	&1	&  0.2002 &   0.0011 &  -0.1510 &  -0.7974\\
		\hline
		\text{identification 3}	&1 &  0.1977  &  0.0046   &-0.1548   &-0.7997  \\
		\hline
		\text{identification 4}	&1 &  0.2120 &  0.0003 &  -0.1485  & -0.8006  \\
		\hline \hline
	\end{tabular}
\vskip 0.2in
	\begin{tabular}{||l|c|c|c|c||} 
		\hline \hline
		\textbf{system} & \textbf{Value 6}& \textbf{Value 7}& \textbf{Value 8}& \textbf{Value 9}\\
		\text{\#} &$b_{1}\,b_{2}$& \text{$\gamma$}& \text{$s^2$}& \text{estimation of $s^2$}\\
		\hline \hline
		\text{true parameters}	 & -1 & - & -&-\\
		\hline\hline
		\text{identification 1}	&-0.9996& 0.2410 &0.1 & 0.1000\\
		\hline
		\text{identification 2}	&  -1.0004& 0.5187 &0.5 & 0.4980\\
		\hline
		\text{identification 3}	& -0.9966& 0.6494 & 1& 1.0010\\
		\hline
		\text{identification 4}	& -1.0017  & 0.8516 & 2& 1.9950\\
		\hline \hline
	\end{tabular}
\end{table*}

\section{Numerical Results} \label{result}
In the following examples,  we address the problem of identifying a two-modes switched system of the form of equation \eqref{eq:ex}, whose true coefficients are   $a_{1}=0.3,~ b_{1}=1,~ a_{2}=-0.5,$ and $b_{2}=-1$. Measurement noise is assumed to be zero-mean with Normal distribution.
In the numerical examples presented, $N=10^{6}$ input-output data is given. 
 True and identified coefficients for different  variances of noise, are presented in Table~\ref{tab:table1}. Variance of  noise and  noise to output ratio for each experiment are also  shown in this table. 
The provided noise to output ratio ($\gamma$) is defined as
\begin{equation}
\gamma = \dfrac{\max \, |\eta|}{\max \, |y|}
\end{equation}

Results are as expected even for high values of noise in comparison to output. As it is illustrated in Table~\ref{tab:table1}, the identified parameters are very close to true values which   demonstrates  the convergence of  proposed algorithm even for small signal to noise ratio.
Moreover, the algorithm requires a very small computational effort. For the case of~$10^6$ measurements and using an off-the-shelf core i5 laptop with 8 Gigs of RAM, the running time  is between 7 to 8 seconds, which shows the effectiveness of approach for very large data sets.

\begin{figure}[h!]
\begin{center}
		\centering\includegraphics[trim={1cm 0.3cm 1cm 1cm},clip,width=0.9\columnwidth]{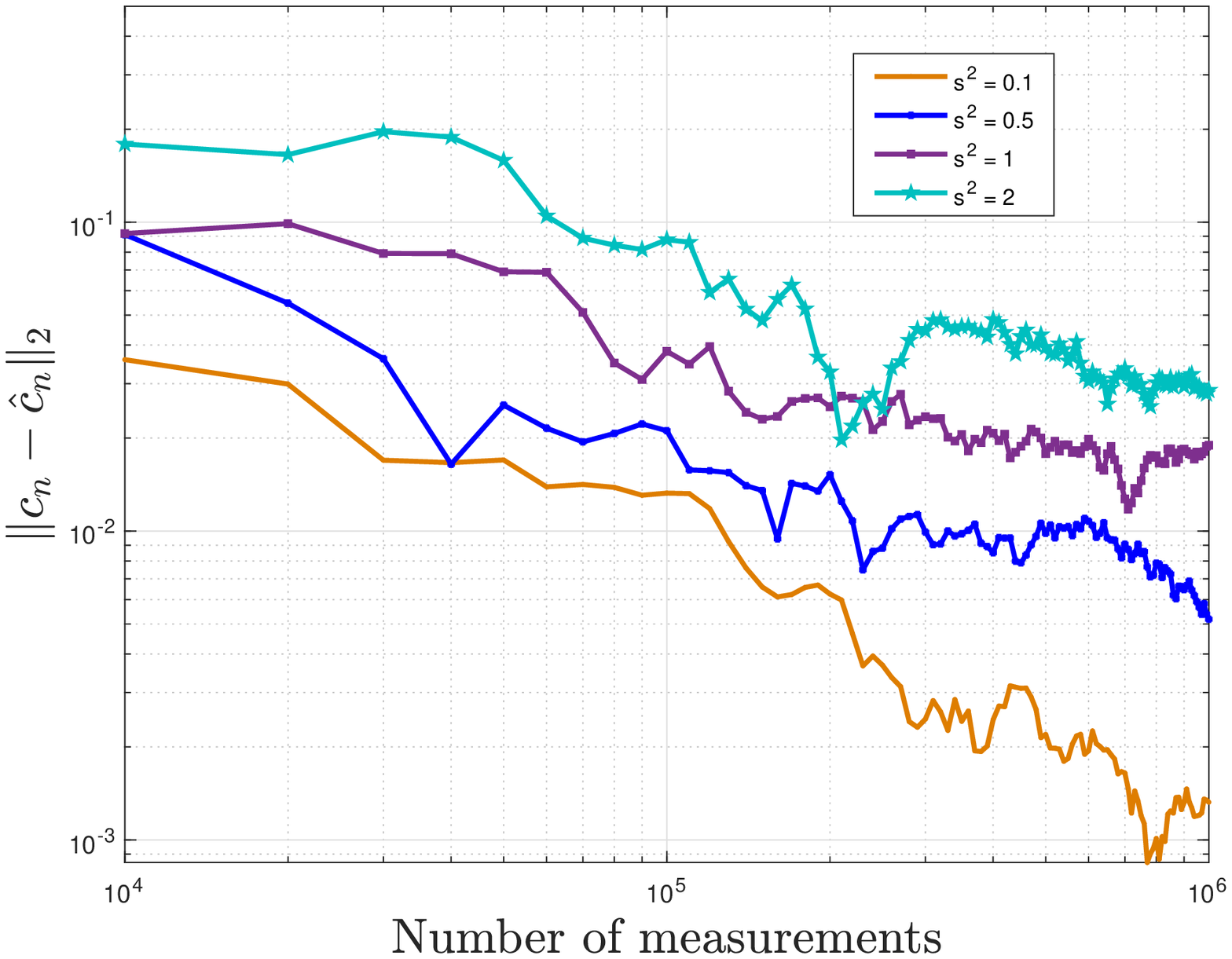}
	\caption{Estimation error of system coefficients}\label{fig:comp}
\end{center}
\end{figure}

The second norm of error between true coefficients of system and estimated coefficients, $\|c_n-\hat{c}_n\|_2\,$, as a function of number of measurements, $N$, is depicted in Figure \ref{fig:comp} for different values of noise variance. As it can be seen from Figure \ref{fig:comp}, the error decreases as the number of measurements increases. Rate of convergence is fast, despite the fact that, in some of the experiments,  a large amount of noise is used. 
It should be noted that these results are for one experiment, and given that this is a realization of a random process, error is not always decreasing.
 For all values of noise variance, error will eventually decrease and the estimated values of coefficients converge to the true values.

Now that we have identified the coefficients of polynomial, it is time to identify each subsystems' coefficients. For the above mentioned example, Table \ref{tab:table2} shows the values of subsystems coefficients for different experiments related to different  values of noise variance. As we see in this table the value of coefficients are very close to the true values, even when the noise variance is high with noise magnitude in average around  85\% of the signal magnitude.

\begin{table}[ht]
	\caption{Identifying submodels' coefficients for different values of noise variance.}
	\label{tab:table2}
	\centering
	\begin{tabular}{l|c|c|c|c|c} 
		\hline \hline
		\text{coeff-} &true&  variance & variance &variance  &variance \\
		\text{icients} & \text{values}& $s^2=0.1$& $s^2=0.5$& $s^2=1$& $s^2=2$\\
		\hline \hline
		$a_{1}$	&   0.3  & 0.3002 &0.2981     &0.3006      &  0.2938    \\
		\hline
		 $b_{1}$	&1	&  0.9988     &  1.0007     & 0.9412      &   1.0031    \\
		\hline
		$a_{2}$	&-0.5	&  -0.4996 &   -0.5000 &  -0.5006 &  -0.5059\\
		\hline
		$b_{2}$	&-1 &  -0.9999  &  -0.9991  & -1.0004&-1.0011 \\
		\hline \hline
	\end{tabular}
\end{table}

The estimation of noise variance based on the structure of matrix $M_{k}$ is shown in Table \ref{tab:table1} as well. The estimates of noise variance are very close to the true values of variance. By knowing the structure of matrix $M_k\,$, the dependence of every entry on the  moments of noise, and the relation in between these moments and the unknown variance (see Section~\ref{sec:notation}), we are able to estimate the noise parameter (in this case, noise variance). This illustrates the capability of the proposed algorithm to estimate both system and noise parameters even for large values of noise.

\begin{figure}[h!]
\begin{center}
	\subfigure[Case 1: true noise variance $s^2= 1$.]{
		\centering\includegraphics[trim={1cm 0.2cm 1cm 1cm},clip,width=0.8\columnwidth]{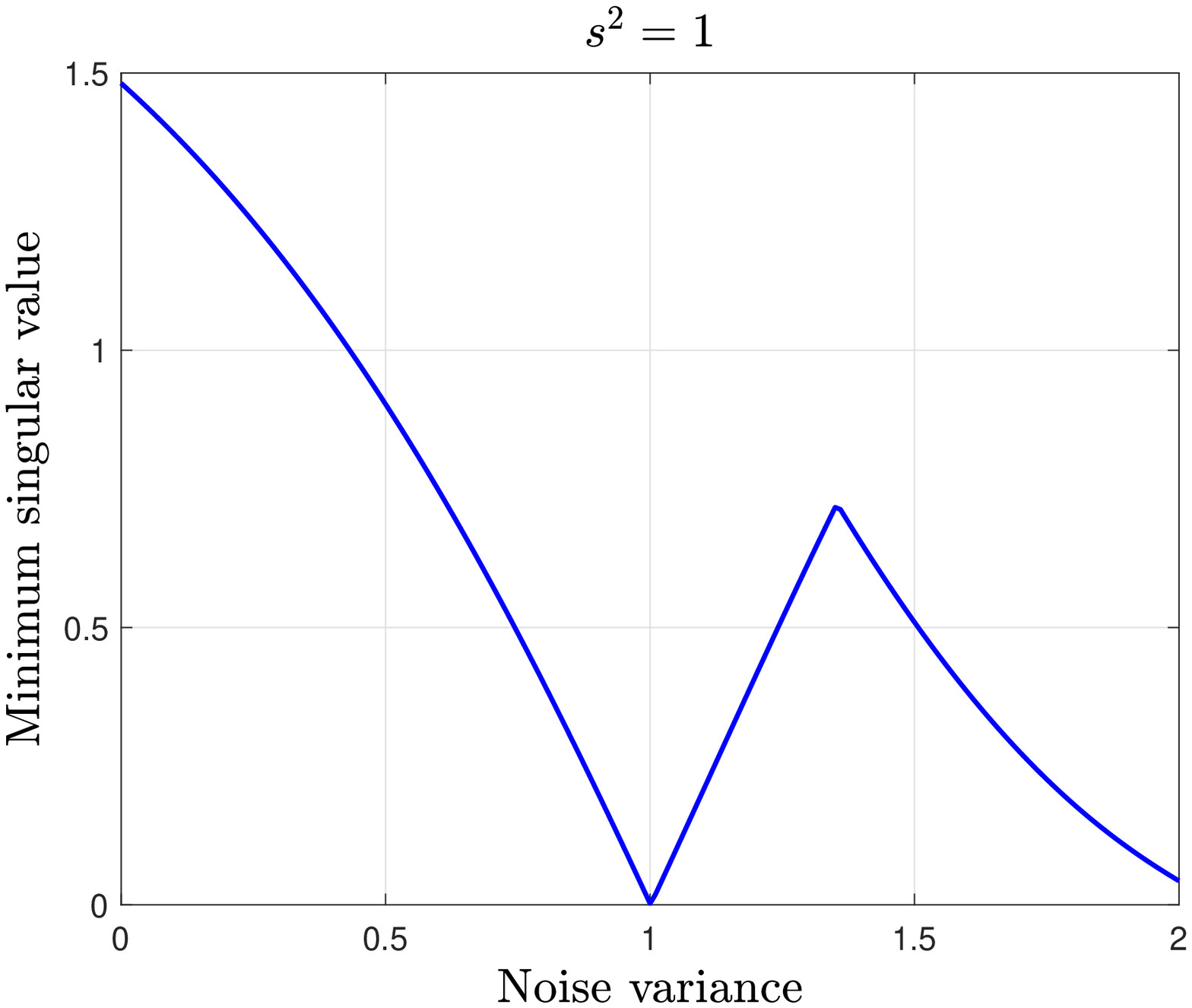} \label{fig:2a}
	}
	\subfigure[Case 2: true noise variance = $s^2=2$.]{
		\centering\includegraphics[trim={1cm 0.3cm 1cm 1cm},clip,width=0.8\columnwidth]{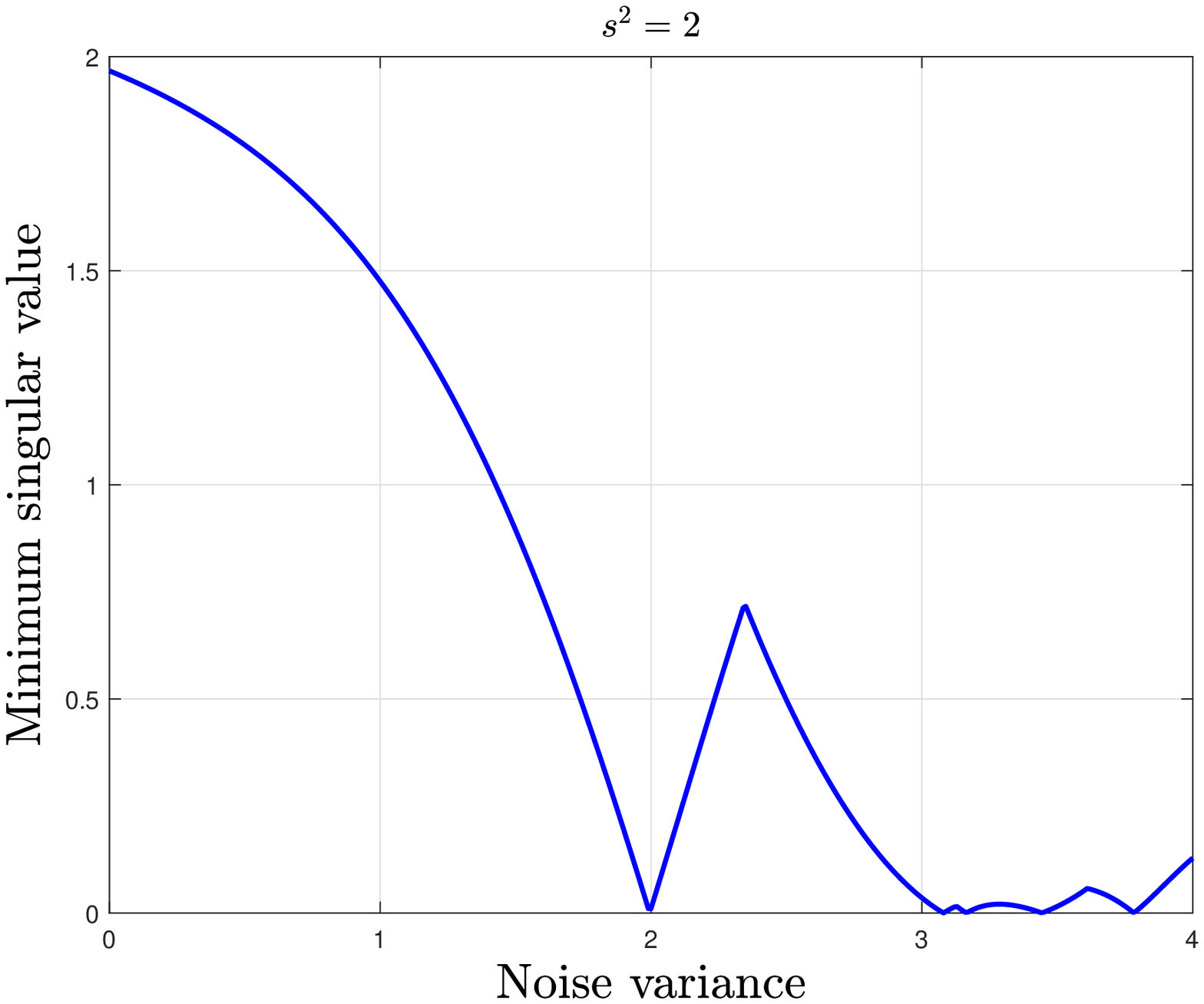} \label{fig:2b}
	}
	\caption{Estimation of noise variance}\label{fig:2}
\end{center}
\end{figure}

 Two examples of the process of estimating the unknown variance of noise are shown in Fig.~\ref{fig:2}; where Fig.~\ref{fig:2a} is for the case of given data contaminated with  noise of variance 1, and Fig.~\ref{fig:2b} is for data with measurement noise of variance 2. By taking $s^*$ as the smallest local minimum, the estimated variance for both cases in Fig.~\ref{fig:2a} and Fig.~\ref{fig:2b} is very close to the true values.

\section{conclusion and future work } \label{conclusion} 
In this paper we have proposed a methodology to identify the coefficients of switched autoregressive  processes and  unknown noise parameters, starting from partial information of the noise and 
given input-output data of switched system.  The approach is shown to be particularly efficient in the case of large amount of data, situation that makes it possible to exploit  law-of-large-numbers type of results. The approach requires the computation of singular value decomposition of a specially constructed input-output Veronese matrix. The ensuing singular vector is then related to the switched system parameters to be identified.
We prove that the estimated parameters converge to the true ones as the number of measurements grows.
Numerical simulations show a low estimation error, even in the case of large measurement noise.
Also, in cases that noise distribution  is not completely known, simulation results show very close estimation of unknown parameters  of noise to the true values. 
In future work, we will consider the problem of identifying switched systems with process noise from large amount of noisy data. Moreover, we will address the problem of identifying switching dynamics in  switched processes form large noisy data sets.
%
\bibliographystyle{plain}
\bibliography{ms}

\appendix
\noindent
\textbf{Sketch of Proof of Theorem~\ref{thm1}:} For simplicity of presentation, let
\[
\widehat{M}_k \doteq M[mon_n(y_k, \ldots, y_{k-n_a})]
\]
We first note that, given the assumptions made on the noise, $u_k$ and $x_k$, the entries of $\widehat{M}_k$ have a variance uniformly bounded for all $k$. Moreover
\[
k>l+n_a \Rightarrow \widehat{M}_k \text{ and } \widehat{M}_l \text{ are independent.}
\]
Hence, by Kolmogorov's Strong  Law of Large Numbers \cite{Sen1993} we have 
\[
\frac{1}{L} \sum_{l=1}^L \widehat{M}_{k+l(n_a+1)} - 	\frac{1}{L} \sum_{l=1}^L E[\widehat{M}_{k+l(n_a+1)}] \rightarrow 0 \text{ a.s.}
\]
as $L \rightarrow \infty$. Since 
\[
E[\widehat{M}_{k)}] = {M}_{k} \text{ for all positive integer } k
\]
and applying the results above for $k=1,2,\ldots,n_a+1$, we conclude that 
\[
\frac{1}{N} \sum_{j=1}^N \widehat{M}_{j} - 	\frac{1}{N} \sum_{j=1}^N M_j \rightarrow 0 \text{ a.s.}
\]
as $N \rightarrow \infty$.
\end{document}